%% file: main.tex
\documentclass[sn-mathphys-ay]{sn-jnl}% Math and Physical Sciences Author Year Reference Style

\graphicspath{{figures/}{pictures/}{images/}{./}} % where to search for the images

\usepackage[table,xcdraw]{xcolor}%
\usepackage{graphicx}%
\usepackage{amsmath,amssymb,amsfonts}%
\usepackage{amsthm}%
\usepackage{mathrsfs}%
\usepackage[title]{appendix}%
\usepackage{xcolor}%
\usepackage{textcomp}%
\usepackage{manyfoot}%
\usepackage{booktabs}%
\usepackage{algorithm}%
\usepackage{algorithmicx}%
\usepackage{algpseudocode}%
\usepackage{listings}%
\usepackage{bm}%
\usepackage{multirow}%
\usepackage{slashbox}%
\usepackage{natbib}
\usepackage{hyperref}

\begin{document}

\title[Article Title]{Do Boxes Affect Exploration Behavior and Performance in Group-in-a-box Layouts?}

%%=============================================================%%
%% GivenName	-> \fnm{Joergen W.}
%% Particle	-> \spfx{van der} -> surname prefix
%% FamilyName	-> \sur{Ploeg}
%% Suffix	-> \sfx{IV}
%% \author*[1,2]{\fnm{Joergen W.} \spfx{van der} \sur{Ploeg} 
%%  \sfx{IV}}\email{iauthor@gmail.com}
%%=============================================================%%

\author*[1]{\fnm{Yuki} \sur{Ueno}}\email{yueno@asu.edu}

\author[2]{\fnm{Hiroaki} \sur{Natsukawa}}\email{natsukawa@g.osaka-seikei.ac.jp}

\author[2]{\fnm{Koji} \sur{Koyamada}}\email{koyamada@g.osaka-seikei.ac.jp}

\affil*[1]{\orgdiv{Ira A. Fulton Schools of Engineering}, \orgname{Arizona State University}, \orgaddress{\state{Arizona}, \country{United States of America}}}

\affil[2]{\orgdiv{Faculty of Data Science}, \orgname{Osaka Seikei University}, \orgaddress{\state{Osaka}, \country{Japan}}}

%%==================================%%
%% Sample for unstructured abstract %%
%%==================================%%

\input{abstract}

\keywords{Human-centered computing, Visu\-al\-iza\-tion, Empirical studies in visualization, Network visualization}

%%\pacs[JEL Classification]{D8, H51}

%%\pacs[MSC Classification]{35A01, 65L10, 65L12, 65L20, 65L70}

\maketitle

\section{Introduction}\label{sec1}
\input{chapter1}

\section{Related Work}\label{sec2}
\input{chapter2}

\section{Methodology}\label{sec3}
\input{chapter3}

\section{Results}\label{sec4}
\input{chapter4}

\section{Discussion}\label{sec5}
\input{chapter5}

\section{Conclusion}\label{sec6}
\input{chapter6} 

\backmatter

\bmhead{Supplementary information}

All supplemental materials are freely available on \href{https://osf.io/cfm68}{OSF}.

\bmhead{Acknowledgements}

This work was supported by JST CREST Grant Number JPMJCR1511, Japan.

\bibliography{sn-bibliography}% common bib file
%% if required, the content of .bbl file can be included here once bbl is generated
%%\input sn-article.bbl

\end{document}

%% file: abstract.tex
\abstract{
    The group-in-a-box (GIB) layout is an efficient graph drawing method designed to visualize the group structure of graphs. The layout communicates group sizes and both within-group and between-group network structures simultaneously. The layout is characterized by its composition of multiple elements, including nodes, edges, and boxes. However, there is limited empirical guidance on how these elements should be combined. In this paper, we measured participants' task performance and eye movements while identifying the group with the largest number of internal edges. We investigated the effect of visualization elements on task performance while controlling the density of internal edges and the box size. The results revealed that the box size in a GIB layout significantly affects the task accuracy either positively or negatively while eye-tracking data suggests that participants focused on internal edges, not the box size. These findings contribute empirical guidance for GIB layout design and lay the groundwork for future research as GIB layout becomes more widely used.
}

%% file: chapter1.tex
Network structures emerge in various contexts such as social networks and phenotype networks in biology. In terms of network visualization, these networks are often considered to be complex \citep{newman2010networks} because they tend to have a community structure, wherein nodes form tight interconnectivity \citep{girvan2002community, newman2004detecting}. For instance, within the X (formerly Twitter) social network, users are depicted as nodes and edges represent mutual mentions. In such a network, users can be divided into sub-groups using the topology or other features of the graph. As the network grows larger or becomes denser, it becomes challenging to analyze due to the increased number of nodes, edges, and groups, leading to clutter. Therefore, an effective network layout for visualizing clustered graphs is imperative.

The group-in-a-box (GIB) layout, depicted in \autoref{GIB_example}, is an efficient graph-drawing method to visualize the group structure of graphs \citep{rodrigues2011group, chaturvedi2014gib, onoue2017optimal}. It consists of multiple elements, including nodes, edges, and boxes. Although some computational experiments \citep{chaturvedi2014gib, onoue2017optimal} and user studies \citep{ueno2019gib, aoyama2019gib} have evaluated GIB variants, designers lack empirical evidence on whether the combination of these visualization elements might aid or hinder how well people interpret the underlying data.

In this paper, we contribute an empirical study exploring and quantifying how visualization elements affect performance. To investigate the effect of multiple visualization elements, we conducted controlled user experiments, using an eye tracker. In a task involving information extraction from a GIB layout, the visualization elements of the graph can determine the accuracy of extraction. However, other visualization elements unrelated to the task may also significantly affect task performance. Specifically, in numerosity comparison tasks, not only the number of elements in a display but also continuous magnitudes, such as element size, can impact task performance \citep{Gebuis2012TheIB}. Based on this background, we investigated how visualization elements in GIB layout affect numerosity comparison process.

In a controlled study with 27 participants, judgments and eye-tracking data were collected to quantify the accuracy of identifying the group with the largest number of internal edges across two conditions: 1) only nodes and edges are displayed, and 2) boxes, nodes, and edges are displayed. The results revealed that both the density of internal edges and the box size in a GIB layout significantly affect task accuracy. Specifically, the box in a GIB layout demonstrated a positive congruency effect. Additionally, while the presence of boxes did not influence exploration behavior, it led to fluctuations in task accuracy.

%% file: chapter2.tex
\subsection{Graph-drawing Method for Group Structure}
Graph drawings excel at simultaneously showing the network topology and attributes. A common approach to visualizing network structure is to use a force-directed layout in which attractive and repulsive forces are balanced to produce an optimal equilibrium \citep{Eades1984AHF}. This force-directed method has been widely used \citep{Fruchterman1991Graph, Kobourov2013Force}. In addition to this method, several approaches have also been developed to achieve better visualization \citep{Koren2003Drawing, Harel2000Multi, Emilio2013Area, Hachul2005Drawing}.

With the recent growing demand for visualizing networks with group structures, several visualization methods have been developed. \citet{Vehlow2017Visualizing} surveyed visualization techniques for group structures in graphs. They classified visualization methods by two-layered taxonomy: vertex group visualization and vertex group structure. In the first layer, there are four categories: visual node attributes, juxtaposed visualization, superimposed visualization, and embedded visualization. The second layer further subdivides the categories based on the distinguishing visual features: overlapping and/or hierarchy.

\begin{figure}[t]
  \centering
  \includegraphics[width=0.8\columnwidth]{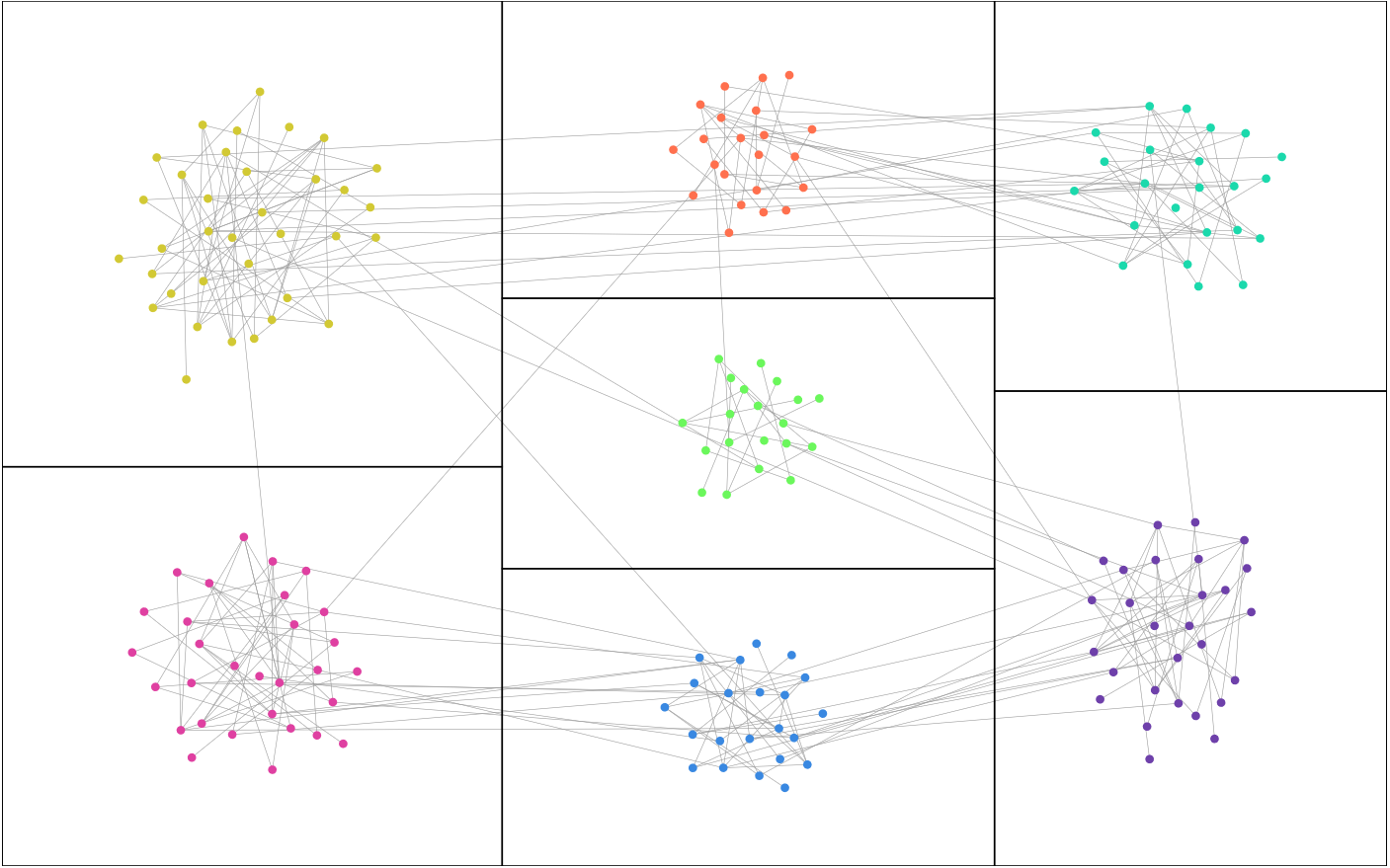}
  \caption{Example of GIB layout. If X (formerly Twitter) data is visualized with the GIB layout, nodes represent users, edges represent messages exchanged between users, and boxes represent communities formed by users who communicate frequently.}
  \label{GIB_example}
\end{figure}

The GIB layout is a graph-drawing method specifically designed to visualize the group structure of graphs \citep{rodrigues2011group, chaturvedi2014gib, onoue2017optimal}. According to Vehlow's taxonomy, this method is categorized as a superimposed visualization and is most suitable for networks without group overlapping. In GIB layouts, all nodes in a group are placed within a box whose size is proportional to the number of nodes. Therefore, by using GIB layouts, it is possible to visualize group structures, group relations, inter-group relations, and the size of the groups simultaneously in a graph. Among several GIB variants, we used Tree-reordered GIB (TR-GIB) (\autoref{GIB_example}).

TR-GIB was developed by \citet{onoue2017optimal}. This layout is based on squarified treemap GIB (ST-GIB) developed by \citet{rodrigues2011group}, which is arranged using the squarified treemap algorithm proposed by \citet{bruls2000squarified}. ST-GIB does not consider the relation between nodes when arranging the boxes; therefore, it includes edge crossings, which tend to hamper users' understanding of the depicted networks \citep{becker1995visualizing, purchase1997aesthetic, purchase1998performance, purchase2002empirical}. TR-GIB is optimized to minimize the lengths of all edges of ST-GIB, resulting in fewer edge crossings compared to ST-GIB. Therefore, this layout offers the advantage of ST-GIB's favorable aspect ratio and effective screen utilization, along with fewer edge crossings.

\subsection{Evaluation of GIB Layout}
Several studies have evaluated GIB layouts. \citet{chaturvedi2014gib} performed computational experiments on three layouts: ST-GIB, force-directed GIB (FD-GIB), and croissant-and-doughnut GIB (CD-GIB). CD-GIB enhances ST-GIB by considering the link information that connects a node to another node in a different group. \citet{onoue2017optimal} demonstrated that TR-GIB is superior to ST-GIB in terms of computational measures. In our previous study, we evaluated ST-GIB, FD-GIB, CD-GIB, and TR-GIB through user experiments where we collected eye-tracking data \citep{aoyama2019gib}. Participants were asked to perform four types of tasks derived from the \citet{Vehlow2017Visualizing}'s task taxonomy for group visualization. The optimal layouts were FD-GIB and TR-GIB, both of which presented various advantages and disadvantages. To elaborate, FD-GIB performed well when grasping aggregate and abstract information about the graph, whereas TR-GIB proved better for displaying specific information about the links in a network.

Furthermore, we conducted an additional controlled experiment to examine exploration behaviors in the task of identifying the group with the largest number of internal edges \citep{ueno2019gib}. This task belongs to group-network tasks, one of the task categories derived from Vehlow's task taxonomy. This task involves all information --group, vertex, and edge information--, thus multiple visual cues may affect task performance. When applied to the X (formerly Twitter) data, this task means identifying the most active community. In the experiment, we kept the number of groups constant and investigated which visualization factors affected the task performance. The results indicated that the exploration behavior was altered by box size, a visualization factor that significantly affected accuracy.

From the results of these experiments, we have identified different kinds of advantages and disadvantages in GIB layouts and the factors that determine performance in GIB layouts. However, the extent to which accuracy is affected by box size remains an open question. By designing and conducting controlled experiments to address this question, we aim to propose guidelines for the effective use of GIB layouts based on human cognition.

%% file: chapter3.tex
To investigate the effect of box size in the task of identifying the box with the largest number of internal edges, we conducted a controlled user experiment.

\subsection{Tasks
\label{task}}
Before the main controlled tasks, participants were given a preliminary task (Task 1) in which they were required to identify the largest box. The goals of this task were to assess how accurately participants could compare box sizes under two different size conditions, which would later be used in the main tasks, and to track their gaze positions while focusing on box sizes. Edges were hidden to prevent overlap with the boxes and to avoid obstructing the side length comparison.

In the main task (Task 2), we measured the effect of the presence or absence of boxes and their size on task performance while identifying the box with the largest number of internal edges. Participants performed under two conditions: without-boxes and with-boxes. This setup allowed us to examine how the presence of boxes influences task performance. Additionally, by controlling box sizes, we assessed how box size affects task performance.

Therefore, we created the following two tasks.
\begin{description}
    \setlength{\itemsep}{0.2cm} % 項目間
    \item[Preliminary Task (Task 1):]Identify the group with the largest number of nodes when only nodes and boxes are displayed.
    \item[Main Task (Task 2):]Identify the group with the largest number of internal edges when boxes are hidden or visible.
\end{description}

\begin{figure*}[t]
    \centering
    \includegraphics[width=\linewidth]{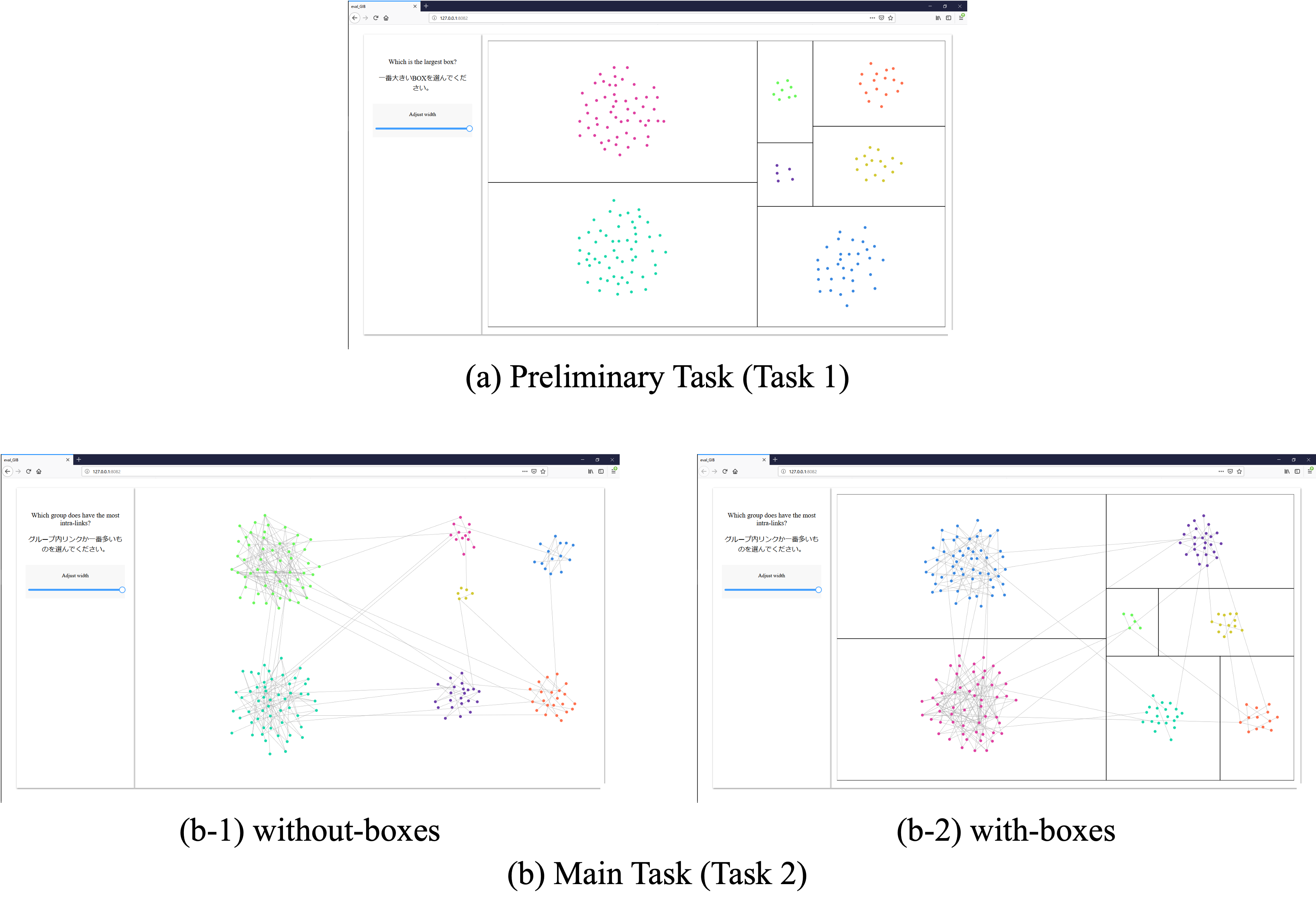} 
    \caption{Examples of each task. (a) Task 1: Identify the group with the largest number of nodes when only nodes and boxes are displayed. (b-1) Task 2 (without-boxes): Identify the group with the largest number of internal edges when only nodes and edges are displayed. (b-2) Task 2 (with-boxes): Identify the group with the largest number of internal edges when the original GIB layout is displayed.}
    \label{task_example}
\end{figure*}

\autoref{task_example} presents an example for each task.

\subsection{Stimuli}
To implement the tasks described in \autoref{task}, visual stimuli were generated. Although the manner of displaying data differed between tasks, the data was generated using the same method to unify the experimental conditions. Each visual element of the GIB layout was set as follows.

\subsubsection{Number of Groups}
Since the experiment aimed not to observe changes in task performance attributable to differences in the number of groups, the total number was consistently fixed at seven. This setting also ensured that there was enough space to display the possible correct answers and to avoid overcomplicating the task of comparing groups.

\subsubsection{Number of Nodes}
In the GIB layout, box size depends on the number of nodes in the group. In our previous work \citep{ueno2019gib}, box size was identified as a factor affecting task performance. Therefore, we devised two scenarios: one where the sizes of the two boxes (the correct answer candidates) were easy to compare, and one where they were difficult to compare. To implement these scenarios, the side ratio ($SR$) of the two correct answer candidates (i.e., the ratio of the side length of the second-largest box to that of the largest box) was set to $0.98$ and $0.91$. These values were chosen to give participants a reasonable chance of correctly identifying the larger box. In the TR-GIB layout, due to its design characteristics, the two rectangular correct answer candidates were positioned vertically with a common long side, so the $SR$ value was applied to the shorter sides.

\begin{figure}[t]
 \centering
  \includegraphics[width=\linewidth]{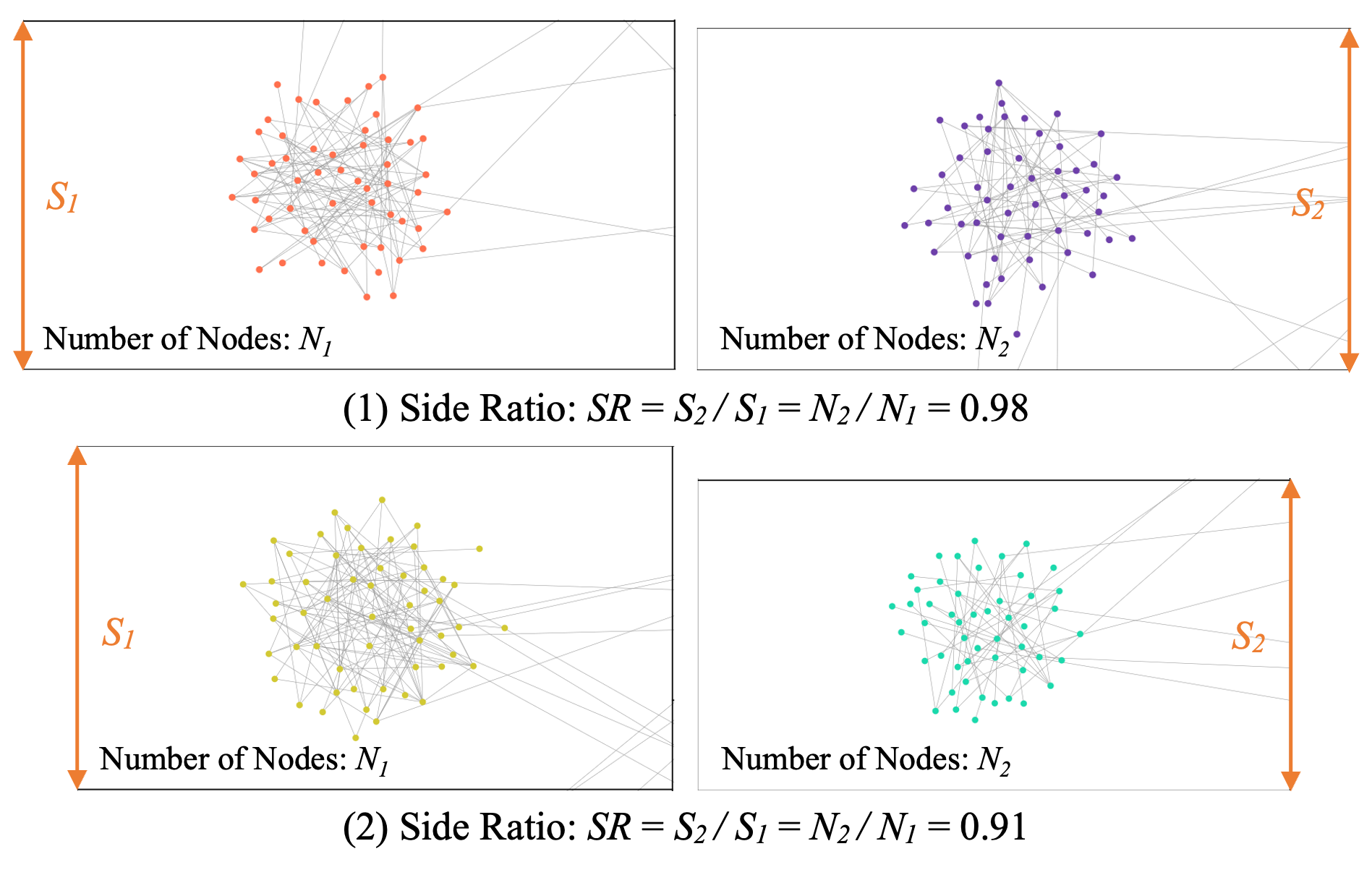} 
 \caption{Example of the side ratio ($SR$) for the correct answer candidates. These boxes share the same long side, while their shorter sides are proportional to the number of nodes.}
 \label{er_explaination}
\end{figure}

The total number of nodes was set at 185, which is the average of the total number of nodes used in the previous study \citep{ueno2019gib}. Let $N_1$ and $N_2$ represent the number of nodes in the two correct answer candidates, and let $N_3$--$N_7$ denote the number of nodes in the other groups. The number of nodes was set as:
\begin{align}
    (N_1, N_2) = \begin{cases}
        (55, 54) & \text{$(SR = 0.98)$} \\
        (55, 50) & \text{$(SR = 0.91)$}
    \end{cases}
\end{align}
\begin{equation}
    N_2-10 \geq N_3,\cdots,N_7 \geq 4 \\
\end{equation}
To prevent the groups with $N_3$ -- $N_7$ nodes from being considered to be the candidate of the correct answer, the $SR$ of these boxes to the second-largest box was set to be smaller than 0.91. \autoref{er_explaination} presents an example of the $SR$ of the correct answer candidates.

\subsubsection{Number of Edges
\label{numofedges}}
During the numerosity comparison process, visual properties are used to estimate numerosity \citep{Gebius2016Sensory, Gevers2016Seosory}. When the number of internal edges is large, participants are likely to select the group with a higher density of internal edges rather than counting and comparing the number of internal edges. Here, the density of internal edges ($D$) is defined as:
\begin{equation}
    D = E/C
\end{equation}
where $C$ is the area of a circular region enclosing all the nodes in a certain group and $E$ is the number of internal edges.

\begin{figure}[t]
    \centering
     \includegraphics[width=\linewidth]{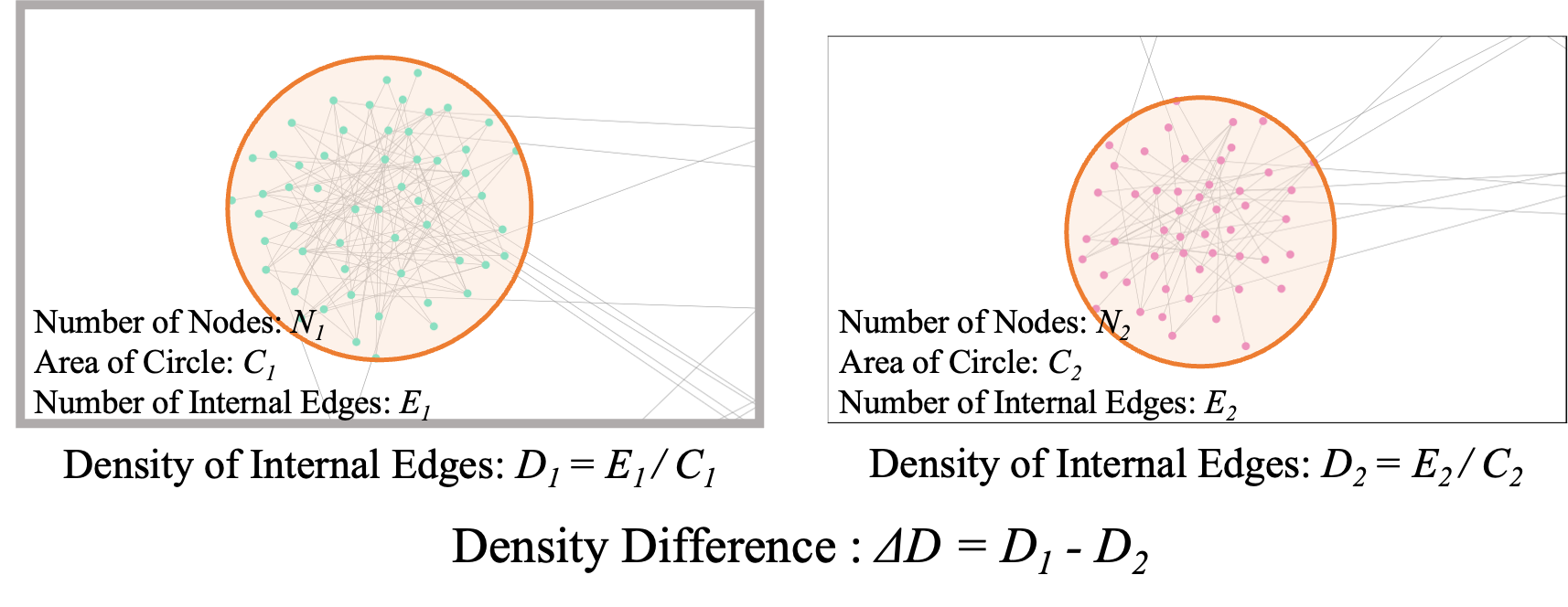} 
    \caption{Explanation of the Density Difference ($\Delta D$). The box on the left contains $N_1$ nodes, an area of $C_1$ for the circle, and $E_1$ internal edges. The box on the right contains $N_2$ nodes, an area of $C_2$ for the circle, and $E_2$ internal edges. The density of internal edges ($D$) is defined as $D_1$ for the left box and $D_2$ for the right box. The density difference ($\Delta D$) is calculated as $D_1 - D_2$. In this example, the left box has a higher density of internal edges than the right box.}
    \label{deltaD_calc_explaination}
\end{figure}

In this experiment, the task difficulty level was controlled by $D$, specifically the difference in $D$ ($\Delta D$) between the two correct answer candidates. $\Delta D$ is defined as:
\begin{equation}
    \Delta D=E_1/C_1-E_2/C_2
\end{equation}
where $E_1$ and $E_2$ are the numbers of edges in the groups with $N_1$ and $N_2$ nodes, respectively. Likewise, $C_1$ and $C_2$ are the areas of the circles enclosing all the nodes in these respective groups (\autoref{deltaD_calc_explaination}).

In our previous research \citep{ueno2019gib}, we generated random data that closely resemble the X (formerly Twitter) data using the method proposed by \citet{chaturvedi2014gib}. As a result, when $\Delta D = 1\times10^{-3}$, accuracy reaches $77.8\%$, and when $\Delta D = -1\times10^{-3}$, accuracy reaches $100\%$. Considering this result, we established the range of $\Delta D$ as follows:

\begin{equation}
    -9\times10^{-4} \leq \Delta D \leq 9\times10^{-4}
\end{equation}

\begin{figure*}[t]
    \centering
    \includegraphics[width=\linewidth]{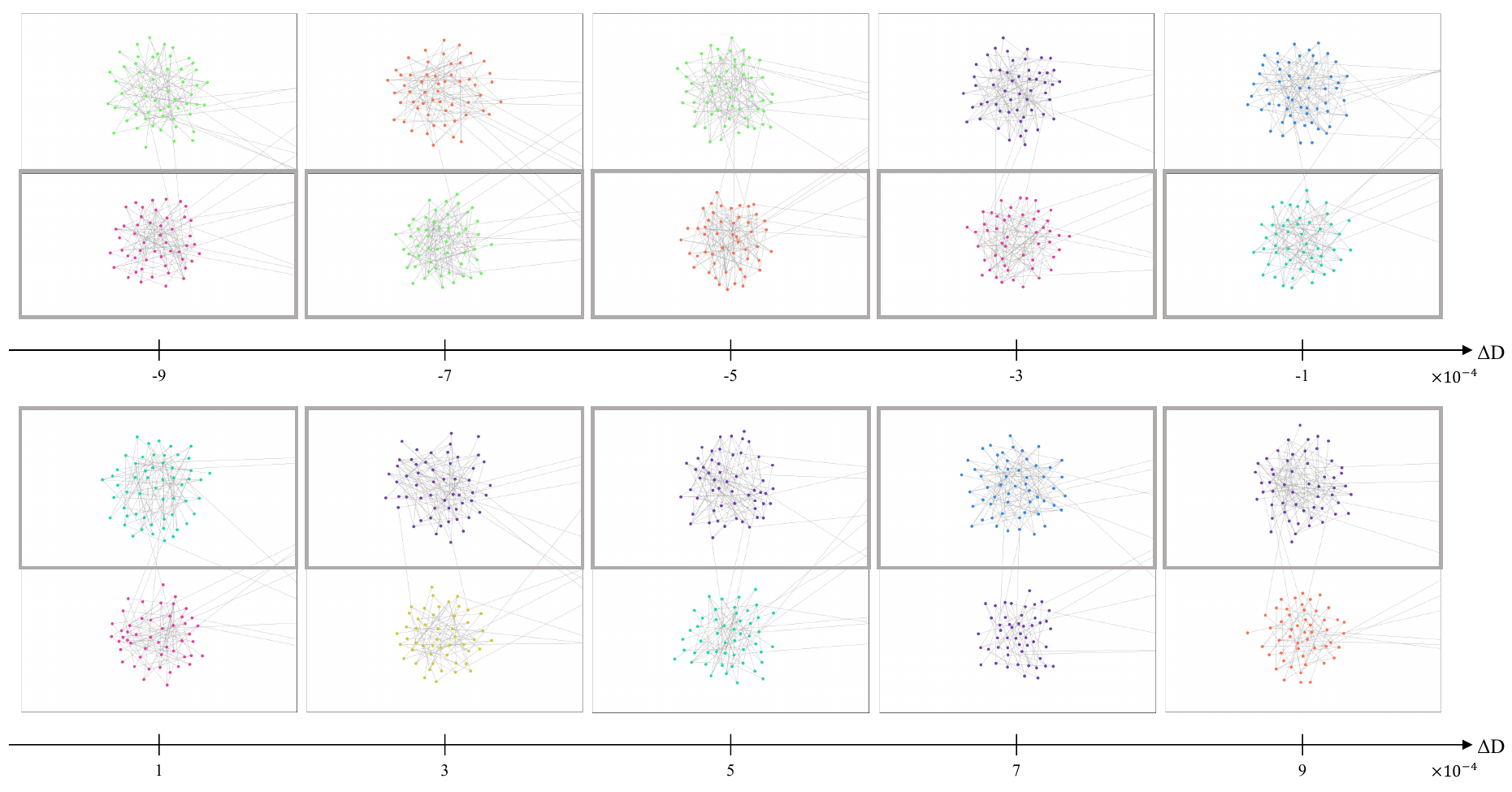} 
    \caption{Examples of the density difference between the groups with $E_1$ and $E_2$ edges when the side ratio is 0.91. The group with more internal edges is marked with a gray border. When $\Delta D$ is less than 0, the group with the second-largest number of nodes ($N_2$) has more internal edges ($E_2$), whereas when $\Delta D$ is greater than 0, the group with the largest number of nodes ($N_1$) has more internal edges ($E_1$).}
    \label{deltaD_explanation}
\end{figure*}

Here, we consider cases where $\Delta D$ is either positive or negative. A positive $\Delta D$ indicates that the group with the largest number of nodes also has the largest number of internal edges. Conversely, a negative $\Delta D$ indicates that the group with the second-largest number of nodes contains the largest number of internal edges. \autoref{deltaD_explanation} provides an example of the density difference between groups with $E_1$ and $E_2$ edges.

In this layout, the circle enclosing all the nodes in each group is determined by the number of nodes, internal edges, and intergroup edges. Consequently, the area of the circle cannot be pre-calculated before generating the visualization. From the data generated with a fixed number of nodes and a variable number of edges, this experiment selected data that met the following conditions:

\begin{description}
    \setlength{\itemsep}{0cm} % 項目間
    \item [Condition 1: $\bm{E_1}$ is the largest number of internal edges]
        \begin{eqnarray}
            \Delta D \approx \{ 1 \times 10^{-4}, 3 \times 10^{-4}, 5 \times 10^{-4},  \nonumber \\
                7 \times 10^{-4}, 9 \times 10^{-4} \} \\
            \bm{E_1} > \bm{E_2} > E_3, \cdots, E_7 \geq 1\\ 
            \bm{E_1/C_1} > \bm{E_2/C_2} > E_3/C_3, \cdots, E_7/C_7\\
            |E'_1/C_1 - E'_2/C_2| \approx 1 \times 10^{-4} 	\label{eq:density}\\
            20 \geq E'_1, E'_2 > E'_3, \cdots, E'_7 \geq 1 \label{eq:intraedge}
        \end{eqnarray}
    \item [Condition 2: $\bm{E_2}$ is the largest number of internal edges]
        \begin{eqnarray}
            \Delta D \approx \{ -1 \times 10^{-4}, -3 \times 10^{-4}, -5 \times 10^{-4}, \nonumber \\
                -7 \times 10^{-4}, -9 \times 10^{-4} \} \\
            \bm{E_2} > \bm{E_1} > E_3, \cdots, E_7 \geq 1\\ 
                \bm{E_2/C_2} > \bm{E_1/C_1} > E_3/C_3, \cdots, E_7/C_7
        \end{eqnarray}
        The other conditions are \autoref{eq:density} and \autoref{eq:intraedge}.
\end{description}
Here, $C_1$ -- $C_7$ represent the areas of the circles that enclose all the nodes in the groups with $N_1$ -- $N_7$ nodes, respectively. $E_1$ -- $E_7$ denote the numbers of internal edges, and $E'_1$ -- $E'_7$ represent the numbers of intergroup edges in each group. The number of intergroup edges ($E'_1$ -- $E'_7$) was limited to a maximum of 20 to prevent excessive intergroup edges, which could detract from the overall readability of the visualization.

\subsubsection{Data Amounts
\label{dataamount}}
We constructed 20 types of datasets based on the combinations of $SR$ values ($0.98$, $0.91$) and $\Delta D$ values ($-9\times10^{-4}$ -- $9\times10^{-4}$). In Task 1, one dataset was generated for each type. In Task 2, five datasets were generated for each type under two conditions (without-boxes and with-boxes), resulting in a total of 220 datasets ($20 \, \text{types} \times (1 + 5 + 5 \, \text{datasets})$).

\subsubsection{Layout}
The generated data were visualized using TR-GIB. This layout can only arrange the boxes; therefore, it is necessary to determine the coordinates of the nodes within each box. Among the available layouts for node placement inside a box, we selected the force layout, which is known for reducing edge crossings and improving readability \citep{Kobourov2013Force}. This method arranges nodes based on the repulsion between nodes, the attraction between adjacent nodes, and the gravity level from the center of the group tile to which they belong. The color scheme for each group was assigned randomly. Although different colors can have varying psychological effects and may influence task performance, this effect was mitigated by randomizing the color assignments.

\subsection{Hypotheses}
We formulated several hypotheses regarding our tasks. Based on the previous study \citep{ueno2019gib}, we observed that box size influenced accuracy. Therefore, we hypothesize that both the size and the presence of the boxes will affect accuracy and exploration behavior.

\begin{description}
    \setlength{\itemsep}{0.2cm} % 項目間
    \item[Hypothesis 1:]In the previous study, accuracy tended to be higher when the correct answer corresponded to the group with the largest area. Therefore, in Task 2 under the with-boxes condition, when $\Delta D > 0$, the accuracy for $SR = 0.91$ exceeds that for $SR = 0.98$, as a lower $SR$ value makes it easier to identify the box with the largest area correctly.
    \item[Hypothesis 2:] In Task 2 under the with-boxes condition, when $\Delta D < 0$, indicating that the correct answer is the group with the second-largest area, the accuracy for $SR = 0.91$ is lower than that for $SR = 0.98$. The contradiction between the box size and the density of internal edges makes it more difficult to select the correct answer.
    \item[Hypothesis 3:]In Task 2 under the with-boxes condition, when $SR = 0.98$, indicating that there is almost no difference in the size of the two potential correct boxes, the presence of the boxes has minimal effect, resulting in no significant difference in accuracy between the with-boxes and without-boxes conditions.
    \item[Hypothesis 4:]In Task 2 under the with-boxes condition, when $SR = 0.91$, indicating a significant difference in the size of the two potential correct boxes, the presence of the boxes has a substantial effect. This results in higher accuracy under the with-boxes condition compared to the without-boxes condition when $\Delta D > 0$, and lower accuracy when $\Delta D < 0$.
\end{description}
Hypotheses 1 to 4 can be tested by evaluating the recorded task accuracy. The following hypothesis can be examined by analyzing the eye-tracking data.
\begin{description}
    \item[Hypothesis 5:]In Task 2, the fixation duration on a graph inside a box is shorter under the with-boxes condition than under the without-boxes condition. Displaying the boxes causes participants to focus on both the graph inside the box and the side lengths of the boxes.
\end{description}

\subsection{Study Design
\label{studydesign}}
As described in \autoref{dataamount}, Task 1 consisted of 20 trials, while Task 2 involved 200 trials, with 100 trials conducted under the without-boxes condition and the other 100 trials under the with-boxes condition. In Task 1, all 20 trials were presented in a single set. In Task 2, the task conditions shifted halfway through: participants first identified the group with the largest number of internal edges under the without-boxes condition, followed by the same task under the with-boxes condition. The 200 trials of Task 2 were organized into eight sets, each comprising 25 trials. The order of both the sets and the individual trials within each task was randomized to minimize potential order effects. Participants took a short break of approximately 30 seconds after completing each set and a longer break, lasting up to 5 minutes, between the two tasks. The eye-tracking system was recalibrated after each break to ensure accuracy.

\subsection{Environmental Conditions and Technical Setup}
The experiment was conducted in our laboratory, which was illuminated by artificial lighting. The tasks were displayed on a 24-inch monitor with a resolution of $1,920 \times 1,080$ pixels. Eye movements were recorded using a Tobii Pro X3-120 eye-tracking system. To minimize head movements during the experiment and reduce noise in the eye-tracking data, participants used a chin rest throughout the tasks.

\subsection{Procedure}
Participants were first given an explanation of the eye-tracking system and the GIB layout. They were then seated 65 cm away from the monitor, as determined by the calibration function of the eye-tracking system. Participants received instructions on the task and performed a tutorial before each task. This training ensured they were sufficiently familiar with the procedure to mitigate the effects of habituation during the actual experiment. The experiment was conducted as described in \autoref{studydesign}. Participants were instructed to perform each task as accurately as possible, with no time constraints imposed. Emphasizing speed could have led to higher error rates and erratic eye movements, which was not the aim of this experiment.

\subsection{Participants}
We employed a within-subject design involving 27 healthy adults, all with normal or corrected-to-normal vision. Among these participants, 21 were male and 6 were female, with an age range of 21 to 33 years and an average age of 24 years. Of the participants, three were familiar with GIB layouts, four had prior experience with visualization, and the remaining 20 had no specialized knowledge of visualization but were experienced in interpreting information from figures and tables.

%% file: chapter4.tex
\subsection{Task Results}
For the statistical analysis, we included the results of all participants. Task accuracy was recorded for each task.

\begin{figure}[t]
 \centering
  \includegraphics[width=0.6\linewidth]{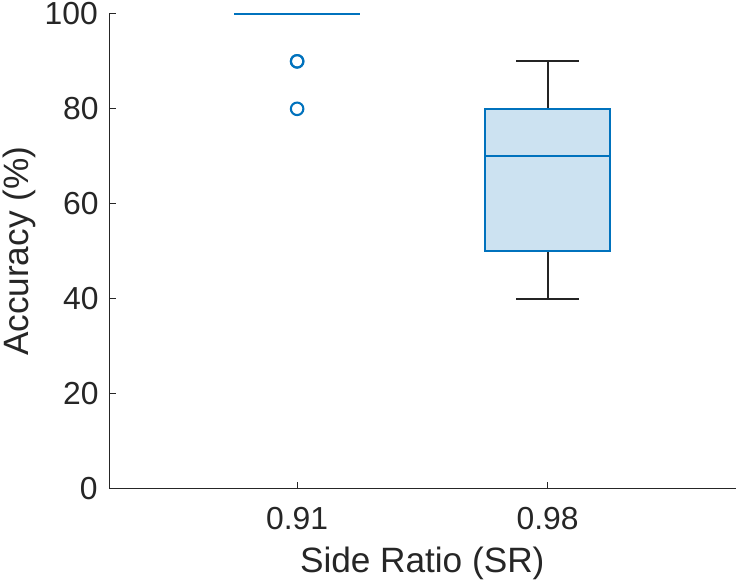}
 \caption{Accuracy results for Task 1. When the box sizes were easy to compare ($SR = 0.91$), the mean accuracy was $97.6\%$. However, when the box sizes were difficult to compare ($SR = 0.98$), the mean accuracy dropped to $66.8\%$.}
 \label{task1}
\end{figure}

\begin{description}
    \setlength{\itemsep}{0.2cm} 
    \item[Tasl 1] In this task, where only nodes and boxes were displayed, participants were asked to identify the group with the largest number of nodes. \autoref{task1} shows the results of this task. The mean accuracy was $97.6\%$ for $SR = 0.91$ and $66.8\%$ for $SR = 0.98$.
    \item[Task 2] In this task, participants were asked to identify the group with the largest number of internal edges. \autoref{task2,3} presents the results. Although we created stimuli under the without-boxes condition using two $SR$ values to standardize the experimental conditions, we merged the results for $SR = 0.98$ and $SR = 0.91$ because participants could not perceive the box size. In this condition (green line), the mean accuracy increased as the absolute value of $\Delta D$ increased. Similarly, under the with-boxes condition (red and blue points), the mean accuracy tends to rise as the absolute value of $\Delta D$ increases.
\end{description}

In addition, we conducted a Wilcoxon signed-rank test to examine whether accuracy differed significantly between the two $SR$ values under the with-boxes condition. The results of this test are presented in \autoref{task3-wilcoxon}. When $\Delta D > 0$, a smaller $SR$ corresponds to higher accuracy, whereas when $\Delta D < 0$, a smaller $SR$ leads to lower accuracy.

\begin{figure}[t]
 \centering
  \includegraphics[width=0.65\linewidth]{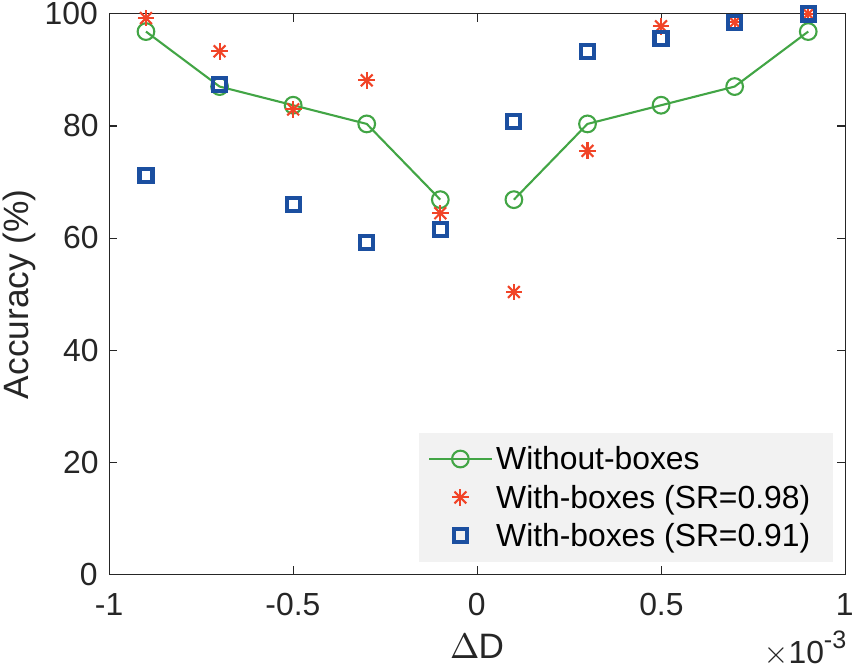}
 \caption{Accuracy results for Task 2. The mean accuracy increased as the absolute value of $\Delta D$ increased. Under the with-boxes condition, at positive $\Delta D$, a smaller $SR$ value resulted in higher accuracy, whereas at negative $\Delta D$, a smaller $SR$ value led to lower accuracy. When comparing the without-boxes and with-boxes conditions, the accuracy under the without-boxes condition was closer to that for $SR = 0.98$ than for $SR = 0.91$ under the with-boxes condition. Additionally, at positive $\Delta D$, accuracy for $SR = 0.91$ under the with-boxes condition tended to be higher than under the without-boxes condition, while at negative $\Delta D$, it tended to be lower.}
 \label{task2,3}
\end{figure}

\begin{table}[tb]
\centering
\caption{Results of the Wilcoxon signed-rank test for the difference in accuracy between the two $SR$ values for each $\Delta D$ in Task 2 under the with-boxes condition. The red highlighted areas indicate where the accuracy for $SR = 0.91$ is significantly greater than for $SR = 0.98$ ($p < 0.05$), while the blue highlighted areas indicate where the accuracy for $SR = 0.91$ is significantly lower than for $SR = 0.98$ ($p < 0.05$).}
\label{task3-wilcoxon}
\begin{tabular}{cccccc}
\toprule
$\Delta D$       & $-9\times10^{-4}$ & $-7\times10^{-4}$ & $-5\times10^{-4}$ & $-3\times10^{-4}$ & $-1\times10^{-4}$ \\
p-value & \cellcolor[HTML]{87CEED}$5.57\times10^{-5}$   & 0.194  & \cellcolor[HTML]{87CEED}$9.46\times10^{-3}$  & \cellcolor[HTML]{87CEED}$4.88\times10^{-4}$  & 0.547  \\ \bottomrule
        &    &    &    &    &    \\ \toprule
$\Delta D$       & $1\times10^{-4}$  & $3\times10^{-4}$  & $5\times10^{-4}$  & $7\times10^{-4}$  & $9\times10^{-4}$  \\
p-value & \cellcolor[HTML]{FFD0BF}$5.50\times10^{-5}$  & \cellcolor[HTML]{FFD0BF}$1.24\times10^{-4}$  & 0.508  & 1.00  & 1.00  \\ \bottomrule
\end{tabular}
\end{table}

\subsection{Eye-tracking Results}
\begin{figure}[t]
 \centering
  \includegraphics[width=0.95\linewidth]{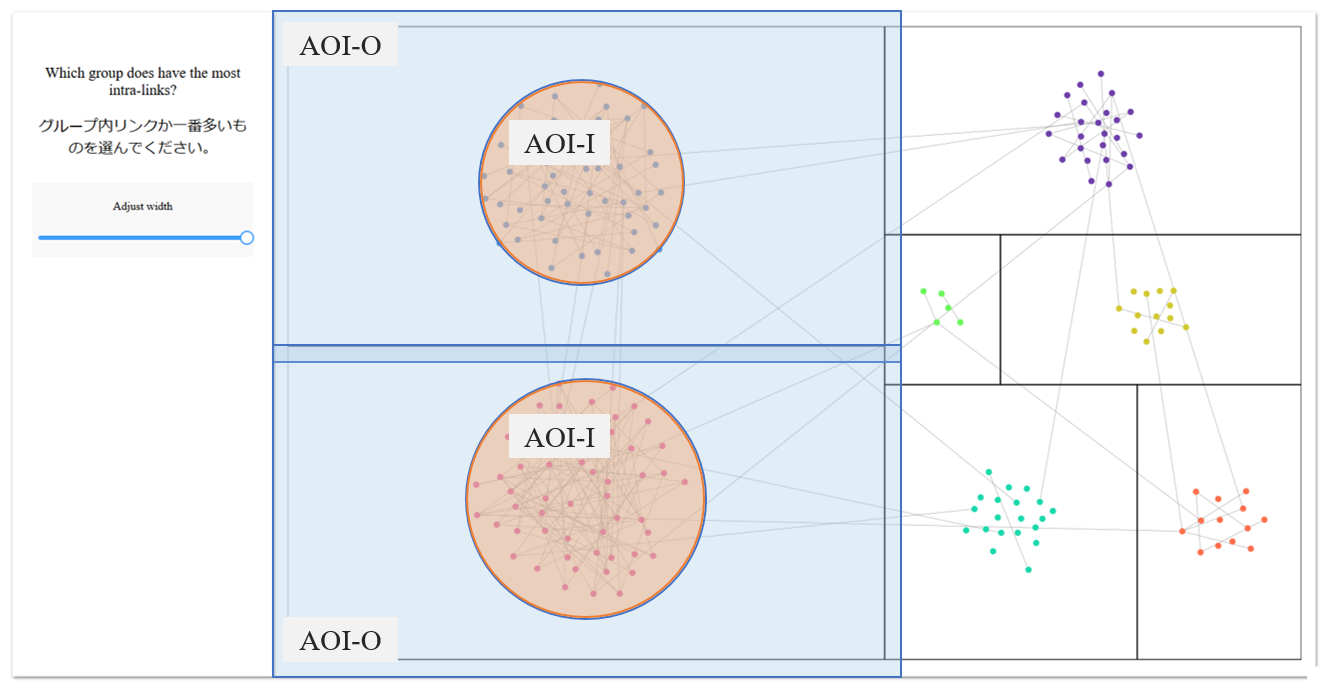}
 \caption{Definition of areas of interest (AOIs). The area of the circle enclosing all the nodes belonging to a specific group is defined as AOI-Inside (AOI-I), while the area slightly larger than the box is defined as AOI-Outside (AOI-O). }
 \label{AOI_explain}
\end{figure}

In addition to task accuracy, eye-tracking data was recorded for each task. The data from six participants were excluded from this analysis due to eye-tracking calibration errors. Our analysis of the eye-tracking data was based on areas of interest (AOIs), as defined in \autoref{AOI_explain}. For the two potential correct boxes, we defined two AOIs: AOI-Inside (AOI-I) and AOI-Outside (AOI-O). AOI-I is a circular area enclosing all the nodes within the group, while AOI-O is a slightly larger area that extends 13.5 pixels beyond the box. We observed that participants' gaze positions tended to be distributed around the periphery of the box when comparing side lengths in Task 1. To accurately capture this comparison behavior as fixations within AOI-O, we slightly expanded the AOI-O area.

The goal of this analysis is to examine whether the presence of boxes affects exploration behavior. If participants actively use the boxes as visual cues, it could alter the fixation duration on AOI-I and AOI-O. Specifically, if participants focus on the boxes, the fixation duration on AOI-O would increase, whereas if they focus on the internal edges, the fixation duration on AOI-I would increase.

For each trial, we calculated the ratio of fixation duration on AOI-I to the sum of fixation durations on both AOI-I and AOI-O, starting 100 ms after the stimulus onset and continuing until the stimulus ended. The average ratio of fixation duration is shown in \autoref{eyetrack-task1,2,3}. In Task 1, participants' gaze was more often directed toward AOI-O, whereas in Task 2, it was primarily focused on AOI-I.

\begin{figure}[t]
 \centering
  \includegraphics[width=0.6\linewidth]{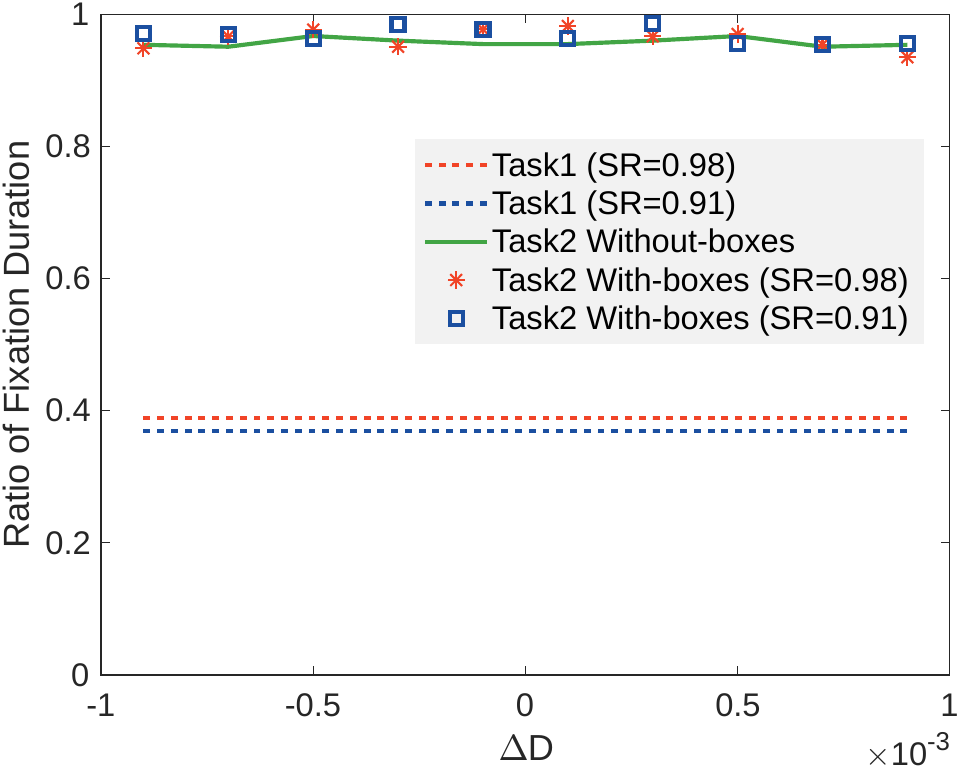}
 \caption{Results of the ratio of fixation duration on AOI-I to the sum of fixation durations on AOI-I and AOI-O for each task.}
 \label{eyetrack-task1,2,3}
\end{figure}

%% file: chapter5.tex
\subsection{Effect of Box Size
\label{box_effect}}
In Task 2, under the with-boxes condition, boxes were displayed, and the $SR$ of the two correct answer candidates was controlled. By comparing the results for $SR = 0.98$ and $SR = 0.91$, we can assess the effect of box size.

According to the results of Task 2 under the with-boxes condition (\autoref{task3-wilcoxon}), there was a significant difference in accuracy between the two $SR$ values at $\Delta D = 1\times10^{-4}$ and $3\times10^{-4}$. Therefore, Hypothesis 1 is supported under these conditions. Although Hypothesis 1 was not fully confirmed, there is a general trend under the with-boxes condition where accuracy for $SR = 0.91$ is higher than for $SR = 0.98$ at positive $\Delta D$.

Additionally, a significant difference in accuracy between the two $SR$ values was observed at $\Delta D = -9\times10^{-4}$, $-5\times10^{-4}$, and $-3\times10^{-4}$. Hence, Hypothesis 2 is supported under these conditions. Although Hypothesis 2 was not entirely confirmed, a general trend regarding the effect of box size was observed. Specifically, accuracy for $SR = 0.91$ was lower than for $SR = 0.98$ at negative $\Delta D$.

When $\Delta D > 0$, this outcome is expected, as for the same $\Delta D$, a larger difference in box size leads to a greater difference in the number of internal edges, resulting in higher task accuracy. Conversely, when $\Delta D < 0$, the correct answer is the second-largest box. In this case, the information about box size contradicts the density information, resulting in lower accuracy compared to when $\Delta D > 0$.

\subsection{Effect of Existence of Box
\label{box_exitst_effect}}
In Task 2, the task was performed under two different conditions: without-boxes and with-boxes. Comparing the results from these conditions allows us to assess the effect of the presence or absence of boxes. The results of the Wilcoxon signed-rank test, which was used to evaluate the statistical difference in accuracy between the without-boxes and with-boxes conditions, are shown in \autoref{task23ttest}.

\begin{table}[tb]
\centering
\caption{Results of the Wilcoxon signed-rank test for the difference in accuracy between the without-boxes and with-boxes conditions in Task 2 for each $\Delta D$. The red highlighted areas indicate where the accuracy under the with-boxes condition is significantly greater than under the without-boxes condition ($p < 0.05$), while the blue highlighted areas indicate where the accuracy under the with-boxes condition is significantly lower than under the without-boxes condition ($p < 0.05$).}
\label{task23ttest}
\begin{tabular}{cccccc}
\toprule
\multicolumn{1}{c|}{\backslashbox{SR}{$\Delta D$}} & $-9\times10^{-4}$ & $-7\times10^{-4}$ & $-5\times10^{-4}$ & $-3\times10^{-4}$ & $-1\times10^{-4}$ \\ \hline
\multicolumn{1}{c|}{0.98} & 0.0625 & \cellcolor[HTML]{FFD0BF}0.0189 & 0.961 & \cellcolor[HTML]{FFD0BF}0.0121 & 0.521 \\
\multicolumn{1}{c|}{0.91} & \cellcolor[HTML]{87CEED}$5.92\times10^{-4}$ & 0.245 & \cellcolor[HTML]{87CEED}$1.17\times10^{-3}$ & \cellcolor[HTML]{87CEED}$3.67\times10^{-4}$ & 0.167 \\ \bottomrule
\multicolumn{1}{l}{}      & \multicolumn{1}{l}{} & \multicolumn{1}{l}{} & \multicolumn{1}{l}{} & \multicolumn{1}{l}{} & \multicolumn{1}{l}{} \\ \toprule
\multicolumn{1}{c|}{\backslashbox{SR}{$\Delta D$}}     & $1\times10^{-4}$ & $3\times10^{-4}$ & $5\times10^{-4}$ & $7\times10^{-4}$ & $9\times10^{-4}$ \\ \hline
\multicolumn{1}{c|}{0.98} & \cellcolor[HTML]{87CEED}$2.83\times10^{-4}$ & 0.285 & \cellcolor[HTML]{FFD0BF}$1.01\times10^{-5}$ & \cellcolor[HTML]{FFD0BF}$1.75\times10^{-5}$ & \cellcolor[HTML]{FFD0BF}0.0313 \\
\multicolumn{1}{c|}{0.91} & \cellcolor[HTML]{FFD0BF}$8.17\times10^{-3}$ & \cellcolor[HTML]{FFD0BF}$1.95\times10^{-4}$ & \cellcolor[HTML]{FFD0BF}$1.28\times10^{-4}$ & \cellcolor[HTML]{FFD0BF}$1.67\times10^{-5}$ & \cellcolor[HTML]{FFD0BF}0.0313 \\ \bottomrule              
\end{tabular}
\end{table}

There was no significant difference in accuracy between the without-boxes and with-boxes conditions for $SR=0.98$ at $\Delta D = -9\times10^{-4}$, $-5\times10^{-4}$, $-1\times10^{-4}$, and $3\times10^{-4}$. Therefore, Hypothesis 3 can be confirmed under these conditions. Although Hypothesis 3 was not fully supported, there is a general trend indicating that the accuracy under the without-boxes condition is closer to that under the with-boxes condition for $SR = 0.98$ than for $SR = 0.91$. This suggests that the effect of the presence of boxes under the with-boxes condition for $SR=0.98$ was smaller than for $SR=0.91$.

The significant difference in accuracy at $\Delta D = 5 \times10^{-4}$, $7\times10^{-4}$, and $9\times10^{-4}$ seems to be due to the effect of box size, as participants were able to correctly discern the slight difference in side lengths 66.8\% of the time, as shown by the results of Task 1. Accuracy under the with-boxes condition for $SR = 0.98$ was higher than under the without-boxes condition when $\Delta D = -7\times10^{-4}$ and $-3\times10^{-4}$, but lower when $\Delta D = 1\times10^{-4}$. This suggests the opposite effect of box size, with visualization elements other than the boxes likely influencing the results.

There was a significant difference in accuracy between the without-boxes and with-boxes conditions for $SR=0.91$, except at $\Delta D = -7\times10^{-4}$ and $-1\times10^{-4}$. Therefore, Hypothesis 4 can be confirmed under these conditions. Although Hypothesis 4 was not fully supported, a general trend regarding the effect of the presence of boxes was observed. Specifically, the accuracy under the with-boxes condition for $SR=0.91$ was higher than under the without-boxes condition at positive $\Delta D$ and lower at negative $\Delta D$.

In psychology, a similar phenomenon is observed in the context of numerical processing. It has been shown that visual properties, such as length, density, and surface area, can impact the ability to compare the numerosity of sets of objects, even when these visual cues are irrelevant to the task \citep{Gebuis2012Con, Gebuis2012TheIB, Gebuis2012TheRV, Judit2020Interplay, Leibovich2017From}. When investigating the process of comparing numerosity, dot comparison tasks are usually used, where two dot arrays are presented and participants are asked to select the dot array with more dots \citep{Gebuis2012TheIB, Leibovich2014Comparing, Judit2020Interplay}. In these tasks, visual properties can be either congruent or incongruent with numerosity. For example, congruent condition means that the array with more dots also has a larger convex hull or a larger dot diameter. In contrast, an incongruent condition means that the array with more dots has a smaller convex hull or a smaller dot diameter. \citet{Gebuis2012TheIB} investigated the effect of visual properties on accuracy in both congruent and incongruent conditions. The results showed that the accuracy significantly differed between congruent and incongruent conditions. For example, accuracy was higher when the convex hull was congruent with numerosity than when it was incongruent. On the other hand, accuracy was lower when the dot diameter was congruent with numerosity than when it was incongruent. Therefore, the convex hull has a positive congruency effect, while the dot diameter has a negative congruency effect.

In this study, when $\Delta D > 0$, the box size is congruent with the number of internal edges. In contrast, when $\Delta D < 0$, the box size is incongruent with the number of internal edges. The results demonstrated that the accuracy was generally higher in the congruent condition than in the incongruent condition. Thus, the effect of the box size has a positive congruency effect. In addition to the box size, other visualization elements are considered to affect the accuracy, which is left for future work.

\subsection{Analysis of Exploration Behavior}
We analyzed how the presence of boxes affected participants' exploration behavior. As shown in \autoref{eyetrack-task1,2,3}, the ratio of fixation duration in Task 1 was about 40\%, indicating that participants focused on the periphery of the boxes when comparing their sizes. In contrast, the ratio of fixation duration in Task 2 was close to 100\%, suggesting a strong tendency to focus on the internal edges. In Task 2, there was no significant difference in the ratio of fixation duration between the without-boxes and with-boxes conditions. Therefore, Hypothesis 5 was not confirmed.

Under the with-boxes condition, we initially expected a lower ratio of fixation duration compared to the without-boxes condition. However, the results of Task 2 showed that participants focused on AOI-I regardless of the presence of boxes. This is consistent with findings on numerosity processing \citep{Li2021Numerosity}, which suggest that peripheral vision can unconsciously influence numerosity judgments by providing contextual information. In our study, peripheral cues from the boxes may have similarly affected participants' accuracy, even when their attention was directed to the graph inside the boxes.

%% file: chapter6.tex
We conducted a user study to investigate the impact of visualization elements on task performance and exploration behavior. Twenty-seven participants were asked to identify the group with the largest number of internal edges while controlling the combination of visualization elements. The results of the study demonstrate that participants focus on the graph within the group, even when the boxes are displayed. Nevertheless, the side length of the boxes can either positively or negatively affect task accuracy. Based on these findings, directly encoding the number of internal edges using visual elements can be suggested as a guideline to mitigate the positive congruency effect of box size. For instance, nodes could be color-coded with a heatmap that reflects the number of internal edges. In future work, the results of this study can serve as a motivation for modeling task accuracy and may provide insights for enhancing the design of GIB layouts.